\documentclass[aps,preprint,a4paper]{revtex4}%
\usepackage{amsmath}
\usepackage{amsfonts}
\usepackage{amssymb}
\usepackage{graphicx}%
\begin{document}
\title{Confinement and stability of the motion of test particles in thick branes}
\author{F. Dahia$^{a}$ and C. Romero$^{b}$}
\affiliation{$^{a}$Dep. de F\'{\i}sica, Univ. Fed. de Campina Grande, 58109-970, Campina
Grande, PB, Brazil}
\affiliation{$^{b}$Dep. de F\'{\i}sica, Univ. Fed. da Para\'{\i}ba, Caixa Postal 5008,
58059,979, Jo\~{a}o Pessoa, PB, Brazil}

\begin{abstract}
{We consider the motion of test particles in thick branes of
`Randall-Sundrum II'-type. It is known that gravity alone cannot
explain the confinement of test particles in this kind of brane.
In this paper we show that a stable confinement in a domain wall
is possible by admitting a direct interaction between test
particles and a scalar field. This interaction is implemented by a
modification of the Lagrangian of the particle which is inspired
by a Yukawa-type interaction between fermions and scalar fields. }

\end{abstract}
\maketitle

Pacs: 04.20.Jb, 11.10.Kk

\section{Introduction}

In braneworld scenarios it is assumed that our observable spacetime
corresponds to a submanifold of a higher-dimensional ambient space. Basically
this picture comes from the idea that matter and non-gravitational fields
might be confined to the ordinary four-dimensional spacetime (the brane) by
some trapping mechanism which would keep extra dimensions hidden from direct
observations at an energy level below TeV scale \cite{Arkani}.

Of course, for the sake of consistency, the existence of extra dimensions
could not be revealed by gravitational probe either at lower energy regime.
Usually this is conceptually achieved by admitting the compactification of
extra dimensions. But, unlike Kaluza-Klein models, extra dimensions could be
much larger than Planck length in braneworld scenarios\cite{Arkani}.

The Randall-Sundrum type II (RSII) model \cite{Randall}, however, surpassing
the compactification paradigm, shows that a zero mode perturbation of gravity
can also be localized in the brane if it is embedded in a non-compact ambient
space such as $AdS_{5}$ (five-dimensional anti-de Sitter) space with $Z_{2}$
symmetry. As a consequence, in the first order approximation, gravity in the
brane is effectively four-dimensional at large distance.

In the RSII model the brane is infinitely thin and possesses a tension which
is responsible for a discontinuity in the first derivative of the metric.
There are solutions involving a real scalar field interacting minimally with
five-dimensional gravity which can be considered as smooth versions of RSII
model \cite{wolfe,chamblin,gremm,Csaki}. The so-called thick branes are
regular domain wall solutions that have a finite thickness and give rise to
well-behaved differentiable metrics. In the infinitely thin limit, these thick
brane solutions reduce to the RSII model \cite{melfo,chico}. Thus, in this
sense, the RSII model might be understood as a kind of idealization of thick
branes scenarios. The confinement of gravity is also valid in thick brane
versions of RSII \cite{wolfe,chamblin,gremm,Csaki}.

In this paper we are interested in studying the motion of test particles in
thick branes focusing on the problem of confinement and stability of the
motion of test particles. If a test particle (a meteorite, in a solar system,
for example) somehow gains a small momentum in the fifth dimension direction,
what happens? Does the particle leave the brane or does it stay moving around it?

The motion of test particle in higher-dimensional spaces has been investigated
under different approaches
\cite{wesson,youm,muck,ishi,seahra,dahia,ponce,anderson,kalil,jalal,romero}.
In Ref. \cite{seahra} necessary general geometric conditions for the geodesic
motion to be stable around a hypersurface were established. It turns out,
however, that the RSII model does not match these conditions. Indeed, it has
been shown that particles, whose geodesic motion is perturbed, will escape
from the brane to infinity in the RSII model \cite{muck}.

As we shall see there is no stable confinement for the thick brane case
either. Then, we can conclude that the gravitational force itself is not
strong enough to confine test particles in this kind of brane. Therefore,
confinement of test particles must be provided by some additional mechanism.

The possibility of localization of matter fields by domain walls was
originally realized by Rubakov and Shaposhnikov \cite{rubakov}. It was shown
that confinement of fermions can be provided by a Yukawa-like interaction
between the spinor Dirac field and a scalar field\cite{rubakov}. Of course,
confinement of test particles must be a consequence of confinement of matter
at the quantum level.

Here we propose a realization of this confinement mechanism in the
non-quantum picture. We shall see that a stable confinement of the
motion of test particle in the thick brane is possible if we
consider a direct interaction between test particles and the
scalar field. Due to this interaction the particles do not follow
a geodesic motion. The interaction with the scalar field is
implemented by a modification of the Lagrangian of test particle
which is inspired by the properties of the Yukawa-type
interaction.

\section{Thick Branes}

Consider a real scalar field $\varphi$ defined in the bulk, which is minimally
coupled with the five-dimensional gravity described by the metric tensor
$\tilde{g}_{AB}.$(Here capital Latin indices run from $0$ to $4$). The total
action of this interaction is%
\begin{equation}
S=\int d^{5}x\sqrt{-\tilde{g}}\left[  -\frac{1}{2\kappa^{2}}\tilde{R}+\frac
{1}{2}\partial_{A}\varphi\partial^{A}\varphi-V\left(  \varphi\right)  \right]
\label{action}%
\end{equation}
where $\tilde{g}$ is the determinant of the metric whose signature is admitted
to be $\left(  -1,+1,+1,+1,+1\right)  $, $\tilde{R}$ represents the scalar
curvature related to $\tilde{g}_{AB}$, $V\left(  \varphi\right)  $ corresponds
to the potential of the scalar field and $\kappa$ is the gravitational
coupling constant in five dimensions.

From that action, we obtain the Einstein equations $G_{AB}=-\kappa^{2}T_{AB}$,
where $T_{AB}$ is the energy-momentum tensor of the scalar field%
\begin{equation}
T_{AB}=\tilde{\nabla}_{A}\varphi\tilde{\nabla}_{B}\varphi-\tilde{g}%
_{AB}\left[  \frac{1}{2}\tilde{\nabla}_{C}\varphi\tilde{\nabla}^{C}%
\varphi+V\left(  \varphi\right)  \right]  \label{T}%
\end{equation}
The dynamic equation for the scalar field is%
\begin{equation}
\tilde{\nabla}_{A}\tilde{\nabla}^{A}\varphi+V^{^{\prime}}\left(
\varphi\right)  =0
\end{equation}
where $\tilde{\nabla}$ is the covariant derivative compatible with $\tilde
{g}_{AB}.$

Let us consider a potential $V\left(  \varphi\right)  $ which has two
degenerate minima at $\varphi=\pm\varphi_{0}$, at least. In the simplest
domain wall solutions, without any further structures, the scalar field
depends only on the extra dimension coordinate, $\varphi=\varphi\left(
y\right)  $. Asymptotically the field tends to the minima values, i.e.,
$\varphi\rightarrow\pm\varphi_{0}$ for $y\rightarrow\pm\infty$.

In order to make contact with the RSII set up\cite{Randall}, let us assume
that the five-dimensional ambient space is a warped product space whose metric
can be written as%
\begin{equation}
ds^{2}=e^{2a(y)}\left(  g_{\mu\nu}dx^{\mu}dx^{\nu}\right)  +dy^{2}
\label{metric}%
\end{equation}
where Greek indices $\mu,\nu=0,1,2,3$ and $g_{\mu\nu}\left(  x\right)  $
represents the induced metric on the hypersurface $y=0$, which is supposed to
be the ordinary four-dimensional spacetime. With no loss of generality we are
assuming here that $a(0)=0$.

Considering the foliation of the ambient space adapted to the Gaussian
coordinate system given above, the Einstein equations can be rewritten in
terms of the intrinsic and the extrinsic curvatures of hypersurfaces $y=const$
$\left(  \Sigma_{y}\right)  $ in the following equivalent form%
\begin{align}
\bar{R}_{\mu\nu}+K_{\mu\nu}K_{\alpha}^{\alpha}-2K_{\mu}^{\alpha}K_{\alpha\nu
}-\frac{\partial K_{\mu\nu}}{\partial y}  &  =-\kappa^{2}(T_{\mu\nu}+\frac
{2}{D-2}\bar{g}_{\mu\nu}T)\\
\bar{\nabla}_{\alpha}K_{\mu}^{\alpha}-\bar{\nabla}_{\mu}K_{\alpha}^{\alpha}
&  =-\kappa^{2}T_{\mu y}\\
-\frac{1}{2}\bar{R}-\frac{1}{2}\left(  K^{2}-K_{\mu\nu}K^{\mu\nu}\right)   &
=-\kappa^{2}T_{yy}%
\end{align}
where $\bar{R}_{\mu\nu}$ and $\bar{\nabla}$ are, respectively, the Ricci
tensor and the covariant derivative associated to the metric $\bar{g}_{\mu\nu
}=e^{2a\left(  y\right)  }g_{\mu\nu}$ induced on the hypersurface $\Sigma
_{y},$ and $K_{\mu\nu}=-\frac{1}{2}\frac{\partial\bar{g}_{\mu\nu}}{\partial
y}$ is the extrinsic curvature of $\Sigma_{y}$. $T=T_{A}^{A}$ is the trace of
the energy-momentum tensor and $D$ is the dimension of the ambient space,
which, in our case, is $D=5$.

The metrics $\bar{g}_{\mu\nu}$ and $g_{\mu\nu}$ (the metric of the ordinary
spacetime) are related by a conformal factor which depends only on the extra
dimension $y$. It follows, then, that $\bar{R}_{\mu\nu}=R_{\mu\nu}$ and
$\bar{R}=e^{-2a\left(  y\right)  }R$.

Under the assumption that $\varphi=\varphi\left(  y\right)  $, the field
equations reduces to a much simpler form. If we admit that the spacetime
metric $g_{\mu\nu}$ is Ricci-flat $\left(  R_{\mu\nu}=0\right)  $, then we
have the following equations%
\begin{align}
a^{\prime\prime}  &  =-\frac{\kappa^{2}}{3}\varphi^{\prime2}\label{a2}\\
a^{\prime2}  &  =\frac{\kappa^{2}}{6}\left(  \frac{1}{2}\varphi^{\prime
2}-V\left(  \varphi\right)  \right)  \label{a1}%
\end{align}
where prime denotes derivative with respect to $y.$

Of course the solution of these equations shall depend on the explicit form of
the function $V\left(  \varphi\right)  $. However, some general aspects of the
behavior of the solutions can be anticipated. A general feature of a domain
wall solution is the existence of a scale parameter $\ell$ which is related to
the width of the thick brane \cite{melfo,chico}. As we have mentioned, in
order to have a domain wall solution, the scalar field $\varphi\left(
y\right)  $ must be a smooth function interpolating between the two real
vacua. Thus, we should have $\varphi\rightarrow$ $\pm\varphi_{0}$ for
$\left\vert y\right\vert >>\ell$ . As a consequence, the equations (\ref{a2})
and (\ref{a1}) asymptotically ($\left\vert y\right\vert >>\ell$) yield%
\begin{align}
a^{\prime\prime} &  =0\\
a^{\prime2} &  =-\frac{\kappa^{2}}{6}V\left(  \pm\varphi_{0}\right)
\end{align}
Assuming that $V\left(  +\varphi_{0}\right)  =V\left(  -\varphi_{0}\right)
<0$, the above equations admit a solution in the form:
\begin{equation}
a(y)=-\sqrt{-\frac{\kappa^{2}}{6}V\left(  \varphi_{0}\right)  }\left\vert
y\right\vert
\end{equation}
which corresponds to the warping factor of the RSII model if we identify
$V\left(  \varphi_{0}\right)  =\Lambda$ (the bulk cosmological constant).

The proper energy density of the domain wall which exceed the vacuum energy
$V\left(  \varphi_{0}\right)  $ is
\begin{equation}
\rho=T_{AB}u^{A}u^{B}-V\left(  \varphi_{0}\right)
\end{equation}
where $u^{A}=e^{-a\left(  y\right)  }\left(  \partial_{t}\right)  ^{A}$ is the
four-velocity of observers which are at rest with respect to the domain wall.
Using (\ref{T}), (\ref{a2}) and (\ref{a1}) and integrating the energy density
in the interval $\eta_{1}\leq y\leq\eta_{2}$ we find%
\begin{equation}
\int_{\eta_{1}}^{\eta_{2}}\left[  \left(  \frac{1}{2}\varphi^{\prime
2}+V\left(  \varphi\right)  \right)  -V\left(  \varphi_{0}\right)  \right]
dy=-\frac{3}{\kappa^{2}}\left(  a^{\prime}\left(  \eta_{2}\right)  -a^{\prime
}\left(  \eta_{1}\right)  \right)  -\int_{\eta_{1}}^{\eta_{2}}\left[  \frac
{6}{\kappa^{2}}a^{\prime2}+V\left(  \varphi_{0}\right)  \right]
dy\label{tension}%
\end{equation}

In the thin brane limit $\left(  \ell\rightarrow0\right)  $, the energy of the
domain wall is supposed to be concentrated around the brane giving rise to the
brane tension $\lambda$%
\begin{equation}
\lambda=\lim_{\ell\rightarrow0}\int_{-\infty}^{+\infty}\left[  \left(
\frac{1}{2}\varphi^{\prime2}+V\left(  \varphi\right)  \right)  -V\left(
\varphi_{0}\right)  \right]  dy
\end{equation}
If the thick brane has a well-defined limit to the RSII model, then the metric
must satisfy some conditions. Let $a_{\ell}\left(  y\right)  $ denote the
warping function related to the parameter $\ell$. In order to guarantee that
the thin brane limit gives the RSII brane, it is necessary that $a_{\ell}(y)$
be locally bounded and $a_{\ell}^{\prime}\left(  y\right)  $ locally
square-integrable for any value of $\ell$ \cite{geroch}. Besides, the sequence
of functions $a_{\ell}(y)$ and $a_{\ell}^{\prime}(y)$ should converge locally
in square integral to $-\sqrt{-\frac{\kappa^{2}}{6}V\left(  \varphi
_{0}\right)  }\left\vert y\right\vert $ and $-\sqrt{-\frac{\kappa^{2}}%
{6}V\left(  \varphi_{0}\right)  }\left(  \theta\left(  y\right)
-\theta\left(  -y\right)  \right)  $ respectively, where $\theta\left(
y\right)  $ is the step function. It turns out then that last integral in
equation (\ref{tension}) goes to zero in the thin brane limit. Thus, we have%

\begin{equation}
\lim_{\ell\rightarrow0}\int_{\eta_{1}}^{\eta_{2}}\left[  \left(  \frac{1}%
{2}\varphi^{\prime2}+V\left(  \varphi\right)  \right)  -V\left(  \varphi
_{0}\right)  \right]  dy=\left\{
\begin{array}
[c]{c}%
\sqrt{-\frac{6}{\kappa^{2}}V\left(  \varphi_{0}\right)  },\qquad\eta
_{1}<0\quad and\quad\eta_{2}>0\\
0,\qquad\eta_{1}<\eta_{2}<0\quad or\quad0<\eta_{1}<\eta_{2}%
\end{array}
\right.
\end{equation}
Therefore, the energy distribution of the thin domain wall is%
\begin{equation}
T_{AB}u^{A}u^{B}=\lambda\delta\left(  y\right)  +V\left(  \varphi_{0}\right)
\end{equation}
with $\lambda=\sqrt{-\frac{6}{\kappa^{2}}V\left(  \varphi_{0}\right)  }$.

If we had considered only the contribution due to the potential we would have
found%
\begin{equation}
V\left(  y\right)  =\frac{\lambda}{2}\delta\left(  y\right)  +V\left(
\varphi_{0}\right)  \label{pot-delta}%
\end{equation}

Let us mention that an example of a regularized version of RSII
brane is the thick brane considered in reference \cite{gremm}. Let
us consider this version here as it leads to the following
analytical solution of the equations
(\ref{a2}) and (\ref{a1}):%
\begin{align}
a\left(  y\right)   &  =-\frac{b\kappa^{2}}{2}\ln\left(  \cosh\left(
\frac{2cy}{\kappa^{2}}\right)  \right)  \label{a}\\
\varphi\left(  y\right)   &  =\sqrt{6b}\arctan\left(  \tanh\left(  \frac
{cy}{\kappa^{2}}\right)  \right)  \label{phi}%
\end{align}
where $b$ and $c$ are positive parameters which are related respectively to
the vacuum value of the scalar field $\left(  \varphi_{0}=\sqrt{6b}%
\pi/4\right)  $ and to the width of the domain wall, respectively. In the
limit $y\rightarrow\pm\infty$, the warping factor converges to
$e^{-2bc\left\vert y\right\vert }$. Thus, in the particular case in which
$g_{\mu\nu}=\eta_{\mu\nu}$, i.e., the Ricci-flat metric $g_{\mu\nu}$ is the
Minkowski metric $\eta_{\mu\nu}$, then (\ref{metric}) corresponds to the
metric of a $Z_{2}$ symmetric $AdS_{5}$ space. Therefore, the thick brane
solution coincides asymptotically with the RSII solution provided the
identification $bc=\sqrt{-\kappa^{2}\Lambda/6}$ is made, where $\Lambda$ is
the bulk cosmological constant. In the thin brane limit, $c\rightarrow\infty$,
with $\left(  bc=\sqrt{-\kappa^{2}\Lambda/6}\right)  $ the RSII solution is
recovered \cite{melfo}.

\section{Confinement of test particles}

By definition test particles carry a very small amount of energy in such way
that they are affected by external fields but the effects they would cause in
the ambient are negligible. As is well known it is assumed that test particles
follow geodesics when they are under the influence of the gravitational field only.

The geodesic motion of a particle with rest mass $m$ can be derived from the
extremization of the action%
\begin{equation}
S=\int m\sqrt{-\tilde{g}_{AB}\dot{x}^{A}\dot{x}^{B}}d\tau\label{S}%
\end{equation}
where the parameter $\tau$ is the particle's proper time and $\dot{x}^{A}$
means $dx^{A}/d\tau$.

Let us consider the motion of a massive test particle in the five-dimensional
warped product space (\ref{metric}). The equation of motion along the extra
dimension $y$ is easily obtained%
\begin{equation}
\,\ddot{y}-e^{2a\left(  y\right)  }a^{\prime}g_{\mu\nu}\dot{x}^{\mu}\dot
{x}^{\nu}=0
\end{equation}
where $a^{\prime}=da/dy$. Using the constraint $\tilde{g}_{AB}\dot{x}^{A}%
\dot{x}^{B}=-1$, we can write%
\begin{equation}
\,\ddot{y}+a^{\prime}(1+\dot{y}^{2})=0\label{ygeo}%
\end{equation}
This shows that the motion of the particle along the extra
dimension decouples from the motion in other directions.
Multiplying equation(\ref{ygeo}) by the integration factor
$e^{2a}\dot{y}$, the following fist integral can be readily
obtained:%
\begin{equation}
e^{2a}\dot{y}^{2}+e^{2a}=\alpha=const.
\end{equation}
Taking $y=0$, we get from the above equation that $\alpha=\dot{y}_{0}^{2}+1$,
where $\dot{y}_{0}^{2}$ might be interpreted as the amount of initial kinetic
energy related to the motion along the extra dimension per rest mass. Thus, we
find%
\begin{equation}
e^{2a}\dot{y}^{2}=\dot{y}_{0}^{2}-\left(  e^{2a}-1\right)  \label{motion}%
\end{equation}
Therefore, given $\dot{y}_{0}^{2},$ the particle can move only in the region
where $\dot{y}_{0}^{2}-\left(  e^{2a}-1\right)  \geq0$. In this sense, we can
say that (\ref{motion}) is equivalent to an equation of motion of a particle
under the influence of the potential $V_{eff}=\left(  e^{2a}-1\right)  $. The
maximum geodesic distance the particle can reach, $y_{\max}$, is then given by
the condition $V_{eff}\left(  y_{\max}\right)  =\dot{y}_{0}^{2}$.

Let us now investigate the motion of a test particle in the thick brane. We
shall see that the motion is not bound to the hypersurface $\Sigma_{0}$.
Indeed, the first derivative of the potential is: $V_{eff}^{\prime}%
=2e^{2a}a^{\prime}$. A necessary condition for the existence of confinement of
test particles in $\Sigma_{0}$ is that $y=0$  be a critical point of $V_{eff}%
$, which, of course, demands that $a^{\prime}\left(  0\right)  =0$.

Thus the function $a\left(  y\right)  $ must satisfy the following `initial
conditions' at $\Sigma_{0}$: $a\left(  0\right)  =0$ and $a^{\prime}\left(
0\right)  =0$. According to equation (\ref{a2}), $a^{\prime\prime}\left(
y\right)  \leq0$ for any $y$, it follows that $a\left(  y\right)  $ has an
absolute maximum at $y=0$. To check this let us suppose that there is a point
$c$ where $a\left(  c\right)  >a\left(  0\right)  =0$. First, let us admit
that $c<0$ . In this case $a^{\prime\,}\left(  m\right)  <0$ for some $m,$
with $c<m<0.$ Then, as $a^{\prime}\left(  0\right)  =0$, $a^{\prime\prime}$
would be positive for some $y$,  $m<y<0$. If $c>0$, then  $a^{\prime\,}\left(
m\right)  >0$ for some $0<m<c.$ Again, $a^{\prime\prime}$ would be positive
for some $0$%
$<$%
$y<m$. Both cases would be in contradiction with equation (\ref{a2}).

Therefore, the function $a\left(  y\right)  $ is non-positive and, as a
consequence, $V_{eff}\leq0$ for any $y$. So, we can conclude that any free
test particle which acquires a non-null transversal velocity $\left(  \dot
{y}_{0}\right)  $ will not stay localized in the vicinity of $\Sigma_{0}$ no
matter how small the velocity is, since its motion has no turning point. This
means that geodesic motion in the brane is unstable. This result is in
accordance with prior analysis which has shown that a particle is expelled
from the thin brane if its motion is perturbed toward the extra dimension
\cite{muck}.

We can illustrate this point by considering the  particular thick brane
described by the solutions (\ref{a}) and (\ref{phi}). In this case, we have
explicitly $V_{eff}=(\left[  \cosh\left(  2cy/\kappa^{2}\right)  \right]
^{-b\kappa^{2}}-1)\leq0$.

It is thus clear that gravity by itself cannot localize test
particles in thick branes of RSII-type. Then, there must exist
some additional mechanism which explains the confinement of them
to the brane. In quantum regime, some mechanisms which play this
role are known. For instance, the Dirac spinor field can be
trapped in a domain wall by a Yukawa-type interaction with the
scalar field \cite{rubakov}. The action for the
five-dimensional massless fermion $\Psi$ which describes this interaction is%
\[
S_{\Psi}=\int d^{4}xdy\left(  i\bar{\Psi}\Gamma^{A}\partial_{A}\Psi
-h\varphi\bar{\Psi}\Psi\right)
\]
where $h$ is the coupling constant. When the classical domain wall solution
$\varphi_{c}\left(  y\right)  ,$ known as kink, is considered then the field
equation for $\Psi$%
\[
i\Gamma^{A}\partial_{A}\Psi-h\varphi_{c}\left(  y\right)  \Psi=0
\]
presents some interesting features. First, the scalar field generates mass for
the five-dimensional fermion. At the scalar field vacua, the fermion acquires
a mass $h\varphi_{0}$. Second, there exists a zero mode solution $\Psi_{0}$
which is localized at $y=0,$ i.e. decays exponentially for large $\left\vert
y\right\vert $ and is proportional to the solution of the usual four
dimensional Dirac field.

The question we face now is how this confinement mechanism can emerge in a
classical picture of test particles.

As we shall see confinement is possible if we admit a direct interaction
between test particles and the scalar field. This interaction is carried out
by a redefinition of the action (\ref{S}). As we have mentioned, in the
quantum picture, the mass of the Dirac field is modified by the scalar field.
Indeed, if we admit that the five-dimensional Dirac spinor field possesses a
rest mass $m,$ then, due to the Yukawa interaction we get the relation%
\begin{equation}
P^{A}P_{A}=-\left(  m^{2}+h^{2}\varphi^{2}\right)  ,\label{energy}%
\end{equation}
where $P_{A}$ represents the $5D$-momentum of the fermion.

Our intention is to define a new Lagrangian in such way that the energy
relation (\ref{energy}) could be derived. One way to do this is changing the
mass term $m$ by $\sqrt{m^{2}+h^{2}\varphi^{2}}$, then the new action for test
particles would read%
\begin{equation}
S=\int\sqrt{m^{2}+h^{2}\varphi^{2}}\sqrt{-\tilde{g}_{AB}\dot{x}^{A}\dot{x}%
^{B}}d\tau\label{newS}%
\end{equation}
Calculating the $5D$-momentum, $\partial L/\partial\dot{x}^{A}$, where $L$ is
the Lagrangian
\[
L=\sqrt{m^{2}+h^{2}\varphi^{2}}\sqrt{-\tilde{g}_{AB}\dot{x}^{A}\dot{x}^{B}}%
\]
we find%
\begin{equation}
P_{A}=-\sqrt{m^{2}+h^{2}\varphi^{2}}\tilde{g}_{AB}\dot{x}^{B}%
\end{equation}
for massive test particles $\left(  \tilde{g}_{AB}\dot{x}^{A}\dot{x}%
^{B}=-1\right)  $. It is easily checked that the condition (\ref{energy}) is
satisfied. Of course, the usual action is obtained again turning off the
interaction, i.e., taking $h=0$. It is worthy of mention that a similar kind
of Lagrangian was also employed, in a different context, to describe the
interaction between test particles and dilatonic fields \cite{kim}.

Let us now investigate the equation of motion of test particles in the warped
product space (\ref{metric}). First, we consider the motion in the extra
dimension. From the Euler-Lagrange equation we get%
\begin{equation}
\sqrt{m^{2}+h^{2}\varphi^{2}}\,\overset{..}{y}+\frac{h^{2}\varphi
\varphi^{\prime}}{\sqrt{m^{2}+h^{2}\varphi^{2}}\,}\dot{y}^{2}+\frac
{h^{2}\varphi\varphi^{\prime}}{\sqrt{m^{2}+h^{2}\varphi^{2}}\,}+\sqrt
{m^{2}+h^{2}\varphi^{2}}a^{\prime}\left(  1+\dot{y}^{2}\right)  =0
\end{equation}
where we have used the constraint $\tilde{g}_{AB}\dot{x}^{A}\dot{x}^{B}=-1$.
Multiplying the equation by the integration factor $\sqrt{m^{2}+h^{2}%
\varphi^{2}}e^{2a}\dot{y},$ the first integral can be immediately obtained and
can be put in the following form%
\begin{equation}
e^{2a}\left(  1+\frac{h^{2}\varphi^{2}}{m^{2}}\right)  \dot{y}^{2}=\dot{y}%
_{0}^{2}-\left[  e^{2a}\left(  1+\frac{h^{2}\varphi^{2}}{m^{2}}\right)
-1\right]  \label{motion-scalar}%
\end{equation}

The effective potential is now $V_{eff}=\left[  e^{2a}\left(
1+\frac {h^{2}\varphi^{2}}{m^{2}}\right)  -1\right]  $, which
reduces to the previous one when $h=0$.

Now let us study the motion of the particle around $\Sigma_{0}$.
As we shall see there are thick branes of RSII-type in which test
particles can be trapped in the domain wall due to the interaction
with the scalar field.

The first derivative of the potential is%
\begin{equation}
V_{eff}^{\prime}=e^{2a}\left[  \left(  2a^{\prime}\right)  \left(
1+\frac{h^{2}\varphi^{2}}{m^{2}}\right)  +2\frac{h^{2}}{m^{2}}\varphi
\varphi^{\prime}\right]
\end{equation}
For the sake of simplicity let us restrict our analysis to the case in which
$\varphi\left(  0\right)  =0$, i.e., the case in which the scalar field is
found in the false vacuum state at $\Sigma_{0}$. Then, $y=0$ will be a
critical point of $V_{eff}$ only if $a^{\prime}\left(  0\right)  =0$.
Calculating the second derivative at $\Sigma_{0}$, we find%
\[
V_{eff}^{\prime\prime}\left(  0\right)  =\left(  2a^{\prime\prime}\left(
0\right)  +2\frac{h^{2}}{m^{2}}\varphi^{\prime2}\left(  0\right)  \right)
\]
Using the field equations (\ref{a2}) and (\ref{a1}) evaluated at $\Sigma_{0}$,
the above expression can be rewritten as%
\[
V_{eff}^{\prime\prime}\left(  0\right)  =4\left(  -\frac{\kappa^{2}}{3}%
+\frac{h^{2}}{m^{2}}\right)  V\left(  0\right)
\]
where $V\left(  0\right)  $ is the value of the potential of the scalar field
at the false vacuum, which is necessarily non-negative according to equation
(\ref{a1}). If $V\left(  0\right)  >0,$ then for the particles that obey the
condition%
\[
m<\frac{\sqrt{3}h}{\kappa}%
\]
the point $y=0$ will represent a stable equilibrium point. Therefore, these
particles will be stably confined to $\Sigma_{0}$ if the energy of
perturbation $\dot{y}_{0}^{2}$ is small enough. In this case, the motion will
be given by the equation $y=y_{0}\sin\omega\tau$ which corresponds to an
oscillation of small amplitude $y_{0}=\dot{y}_{0}/\omega$ and angular
frequency $\omega=\sqrt{2\left(  \frac{h^{2}}{m^{2}}-\frac{\kappa^{2}}%
{3}\right)  V\left(  0\right)  }$. Moreover, if $\omega$ is high enough and
$\dot{y}_{0}$ is sufficiently small, then the period of oscillation will be so
short and the amplitude so small that the movement in the extra dimension
could not be detected by $4D-$observers.

It is clear that the confinement is stronger in cases of large values of
$V\left(  0\right)  $. The width of the brane also influence the intensity of
the binding of the particle to $\Sigma_{0}$. In the previous section, we have
seen that in the thin limit $V\left(  y\right)  $ tends to a delta function
whose support is on $y=0$. Thus, $V\left(  0\right)  $ cannot be bounded in
the limit $\ell\rightarrow0$. Indeed, consider the inequality%
\[
\int_{a}^{b}\left(  V\left(  y\right)  -V\left(  \varphi_{0}\right)  \right)
dy\leq\left[  \max_{y\in\left[  a,b\right]  }\left(  V\left(  y\right)
-V\left(  \varphi_{0}\right)  \right)  \right]  \left(  b-a\right)
\]
Assuming that $V\left(  \varphi\right)  $ has an absolute maximum at
$\varphi=0$ in the range $\left[  -\varphi_{0},\varphi_{0}\right]  $, then%
\[
V\left(  0\right)  -V\left(  \varphi_{0}\right)  \geq\frac{1}{b-a}\int_{a}%
^{b}\left(  V\left(  y\right)  -V\left(  \varphi_{0}\right)  \right)  dy
\]

Consider now the limit $\ell\rightarrow0$ keeping $\Lambda$ constant. For
$a<0$ and $b>0$, according to equation (\ref{pot-delta}), the integral has a
finite value which is independent of $b$ and $a$%
\[
\left(  \lim_{\ell\rightarrow0}V\left(  0\right)  \right)  -V\left(
\varphi_{0}\right)  \geq\frac{1}{2}\frac{\lambda}{b-a}%
\]
But as $b$ and $a$ are arbitrary, the difference $b-a$ can be made
arbitrarily small. Therefore, in the thin limit $V\left(  0\right)
$ must increases with no bounds.

It is worthy of mention that for a particle whose rest mass is
greater than the critical value $\sqrt{3}h/\kappa$, the
interaction is not strong enough to keep it bound to the
hypersurface $\Sigma_{0}$.

Once more let us illustrate this point by considering the particular thick
brane given by  solutions (\ref{a}) and (\ref{phi}). In this case, the
effective potential is%
\begin{equation}
V_{eff}=\frac{1+\frac{6bh^{2}}{m^{2}}\left[  \arctan\left(  \tanh\left(
\frac{cy}{\kappa^{2}}\right)  \right)  \right]  ^{2}}{\left[  \cosh\left(
\frac{2cy}{\kappa^{2}}\right)  \right]  ^{b\kappa^{2}}}-1\label{potential}%
\end{equation}

It can be directly checked that $V_{eff}^{\prime}\left(  0\right)  =0$ and
that $V_{eff}^{\prime\prime}\left(  0\right)  =\left(  3\frac{h^{2}}%
{m^{2}\kappa^{2}}-1\right)  \frac{4bc^{2}}{\kappa^{2}}$. As we have
anticipated in the general case,  $y=0$ is a stable equilibrium point for
particles that obey the condition $m<\sqrt{3}h/\kappa$ (see Fig.1). It can be
easily verified also that the confinement is stronger in cases of small domain
wall width $(c\rightarrow\infty)$. Another interesting characteristic is that
the intensity of the binding is greater in the case of large values of
$\Lambda$ $($recall that $bc=\sqrt{-\kappa^{2}\Lambda/6})$. It is important to
stress that the latter feature is valid to this particular thick brane, but we
cannot assure its general validity.

In this particular thick brane, as the potential is explicitly known,  general
motions, i.e., those not restricted to small perturbations, can be studied.
The potential is an even function which vanishes at $y=0$ and tends to $-1$
for $\left\vert y\right\vert \rightarrow\infty$. If the condition $m<\sqrt
{3}h/\kappa$ holds, $V_{eff}\left(  y\right)  $ has a maximum value $V^{\ast}$
at two points $\pm y^{\ast}$. Therefore the potential is a bounded function:
$-1\leq V_{eff}\leq V^{\ast}.$ A rough estimate of an upper limit of $V^{\ast
}$ can be readily obtained by taking the maximum value of numerator and the
minimum value of the denominator simultaneously in (\ref{potential}). This
gives $V^{\ast}<3bh^{2}\pi^{2}/8m^{2}$. Then, the initial kinetic energy
$\dot{y}_{0}^{2}$ of the light particles ($m<\sqrt{3}h/\kappa)$ cannot exceed
this amount of energy in order that we have confinement, otherwise they
certainly will escape from the brane and will never return.

In the case of confinement, $\dot{y}_{0}^{2}<V^{\ast},$ the
particle will oscillate around the hypersurface $\Sigma_{0}$ and
will reach a maximum distance $y_{\max}$ which is given by

{\small
\begin{equation}
\frac{1+\frac{6bh^{2}}{m^{2}}\left[  \arctan\left(  \tanh\left(
\frac{cy_{\max}}{\kappa^{2}}\right)  \right)  \right]  ^{2}}{\left[
\cosh\left(  \frac{2cy_{\max}}{\kappa^{2}}\right)  \right]  ^{b\kappa^{2}}%
}-1=\dot{y}_{0}^{2}%
\end{equation}
}

Let us define $\widetilde{y}_{max}=cy_{max}$. In terms of this new variable
the above equation does not depend on the parameter $c$. Then, it is clear
that
\begin{equation}
y_{\max}=\frac{\widetilde{y}_{\max}}{c}%
\end{equation}

It is interesting to note that the particle will stay closer to
the hypersurface $\Sigma_{0}$, or in other words, confinement will
be stronger in case of thinner domains wall $\left(
c\rightarrow\infty\right) $.

Now let us consider the general case again. Let us now study the motion of
massless spinor particles by admitting that their motion in the fifth
direction is given by Eq.(\ref{motion-scalar}) in the limit $m\rightarrow0$.
Multiplying that equation by $m$ and taking the mentioned limit, we get%
\begin{equation}
e^{2a}h^{2}\varphi^{2}\dot{y}^{2}=-e^{2a}\left(  h^{2}\varphi^{2}\right)
\end{equation}
The unique solution for this equation is $y=0$. As consequence the energy
relation (\ref{energy}) reduces to%
\begin{equation}
p_{\mu}p^{\mu}=0
\end{equation}
This means that the particle, which seems to be massless with
respect to $4D$-observers, stays strictly confined to the
hypersurface $\Sigma_{0}$. This result can be confirmed by
analyzing the second derivative of the effective potential, which
gives the strength of the binding force between the particle and
the domain wall. As we have seen, it increases for particles with
small mass. Then, it is expected that in the massless limit the
binding force becomes so huge that the particles cannot escape
from $\Sigma_{0}$.

\section{Motion of test particles in a Schwarzschild thick brane}

Let us now consider the issue we have already raised concerning the motion of
a small meteorite (considered as a test particle) in the solar system in the
context of thick branes. For the sake of simplicity, let us admit that the
geometry around the sun is approximately given by the Schwarzschild metric.
Thus, the five-dimensional metric is \cite{hawking}%
\begin{equation}
ds^{2}=e^{2a\left(  y\right)  }\left[  -f\left(  r\right)  dt^{2}%
+f^{-1}\left(  r\right)  dr^{2}+r^{2}d\theta^{2}+r^{2}\sin^{2}\theta d\phi
^{2}\right]  +dy^{2}\label{Schawrzbrane}%
\end{equation}
where $f\left(  r\right)  =\left(  1-\frac{2M}{r}\right)  $.

Since the Schwarzschild metric is Ricci-flat, the field equations
reduce to the same equations (\ref{a2}) and (\ref{a1}). Therefore
the general analysis of the previous section concerning the motion
of test particles in the fifth dimension, which is described by
equation (\ref{motion-scalar}), is also applicable to the case of
the meteorite in the Schwarzschild thick brane
(\ref{Schawrzbrane}). Then we can conclude that, for a small
perturbation, the meteorite stays bound to the hypersurface
$\Sigma_{0}$ and its motion in the fifth dimension might not be
directly observed.

We now study the motion in the other directions. First, let us obtain the
Euler-Lagrange equation for the coordinate $\theta$%
\begin{equation}
\frac{d}{d\tau}\left(  \sqrt{m^{2}+h^{2}\varphi^{2}\left(  y\right)
}e^{2a\left(  y\right)  }r^{2}\dot{\theta}\right)  =r^{2}\sin\theta\cos
\theta\dot{\phi}^{2}%
\end{equation}
As usual we consider the particular solution $\theta=\pi/2$, i.e., the motion
in the equatorial plane.

The equations for $t$ and $\phi$ are%
\begin{align}
\sqrt{m^{2}+h^{2}\varphi^{2}\left(  y\right)  }e^{2a\left(  y\right)
}f\left(  r\right)  \dot{t}  &  =E=const\label{t}\\
\sqrt{m^{2}+h^{2}\varphi^{2}\left(  y\right)  }e^{2a\left(  y\right)  }%
r^{2}\dot{\phi}  &  =L=const \label{angle}%
\end{align}
where $E$ and $L$ are respectively the energy and the angular momentum
measured by an asymptotic observer $\left(  r\rightarrow\infty\right)  $ in
the brane $\left(  y=0\right)  $.

The Euler-Lagrange equation related to the coordinate $r$ can be readily
integrated and, using (\ref{motion-scalar}), (\ref{t}) and (\ref{angle}),
gives%
\begin{equation}
e^{2a}f^{-1}\left(  r\right)  \dot{r}^{2}+\left(  \frac{-E^{2}}{f\left(
r\right)  }+\frac{L^{2}}{r^{2}}\right)  \frac{e^{-2a}}{\left(  m^{2}%
+h^{2}\varphi^{2}\right)  }+\left(  \dot{y}_{0}^{2}+1\right)  \frac
{m^{2}e^{-2a}}{\left(  m^{2}+h^{2}\varphi^{2}\right)  }=0 \label{r}%
\end{equation}
Note that if there is no perturbation, i.e., $\dot{y}_{0}^{2}=0$, then $y=0$
is a solution of (\ref{motion-scalar}) and the equations (\ref{t}),
(\ref{angle}) and (\ref{r}) give the usual four-dimensional motion of a test
particle in the Schwarzschild geometry.

Now let us consider the equation of motion parameterized by the time measured
by an asymptotic observer%
\begin{equation}
\frac{E^{2}}{f^{3}\left(  r\right)  }\left(  \frac{dr}{dt}\right)  ^{2}%
+\frac{-E^{2}}{f\left(  r\right)  }+\frac{L^{2}}{r^{2}}+m^{2}\left(  1+\dot
{y}_{0}^{2}\right)  =0 \label{rt}%
\end{equation}
It is interesting to note that compared to the usual corresponding equation in
the four-dimensional Schwarzschild spacetime the Eq.(\ref{rt}) differs only by
the additional term $m^{2}\dot{y}_{0}^{2}$. Therefore, if the perturbation is
truly small, i.e., $\dot{y}_{0}^{2}<<1$, then, despite the oscillatory motion
in the fifth dimension, the asymptotic observer cannot detect the existence of
a fifth dimension by examining the time evolution $r\left(  t\right)  $ of the meteorite.

The same conclusion is valid for the trajectory, i.e. $r=r\left(  \phi\right)
$. Substituting $\dot{r}=\frac{dr}{d\phi}\dot{\phi}$ in the equation
(\ref{rt}) we have%
\begin{equation}
\frac{L^{2}}{f\left(  r\right)  r^{4}}\left(  \frac{dr}{d\phi}\right)
^{2}+\frac{-E^{2}}{f\left(  r\right)  }+\frac{L^{2}}{r^{2}}+m^{2}(1+\dot
{y}_{0}^{2})=0
\end{equation}
It is remarkable that, as in the previous case, the only modification of the
trajectory equation, due to the fifth dimension, is again the term $m^{2}%
\dot{y}_{0}^{2}$. Thus, the influence of the extra dimension in the trajectory
of the meteorite might be neglected if the energy of perturbation is small enough.

\section{Conclusion}

We have studied the motion of test particles in thick branes of
RSII-type. We have seen that a particle following a geodesic
motion does not stay bound to the four-dimensional spacetime if
its motion is transversally perturbed. This means that gravity
alone cannot be responsible for the confinement of test particles
to this kind of brane.

Based on the mechanism of localization of fermions in a domain
wall by means of a Yukawa-like interaction between fermions and
the scalar field, we have proposed a direct interaction between
test particles and the scalar field by changing the Lagrangian of
test particles.

We have shown that due to this interaction light test particles, i.e.
particles which obey the condition $m<\sqrt{3}h/\kappa$ will be bound to the
brane for small transversal perturbations. Heavier particles if transversally
perturbed will acquire so large an inertia that the interaction with scalar
field will not be strong enough to stop their motion into the extra dimension.

We also analyze the motion of test particles in the Schwarzschild thick brane.
As expected for the consistency reasons of the brane world scenario, in the
lower energy regime the motion of test particles does not reveal the existence
of an extra dimension, neither by direct observation of the oscillatory motion
in the fifth dimension nor by deviation of the motion in the four-dimensional
ordinary spacetime.

\section{Acknowledgement}

The authors thank CNPq-FAPESQ (PRONEX) for financial support. Thanks also go
to the referee for useful comments and suggestions.

%

\begin{figure}
[ptb]
\begin{center}
\includegraphics[
height=2.9603in, width=3.557in, angle=270
]%
{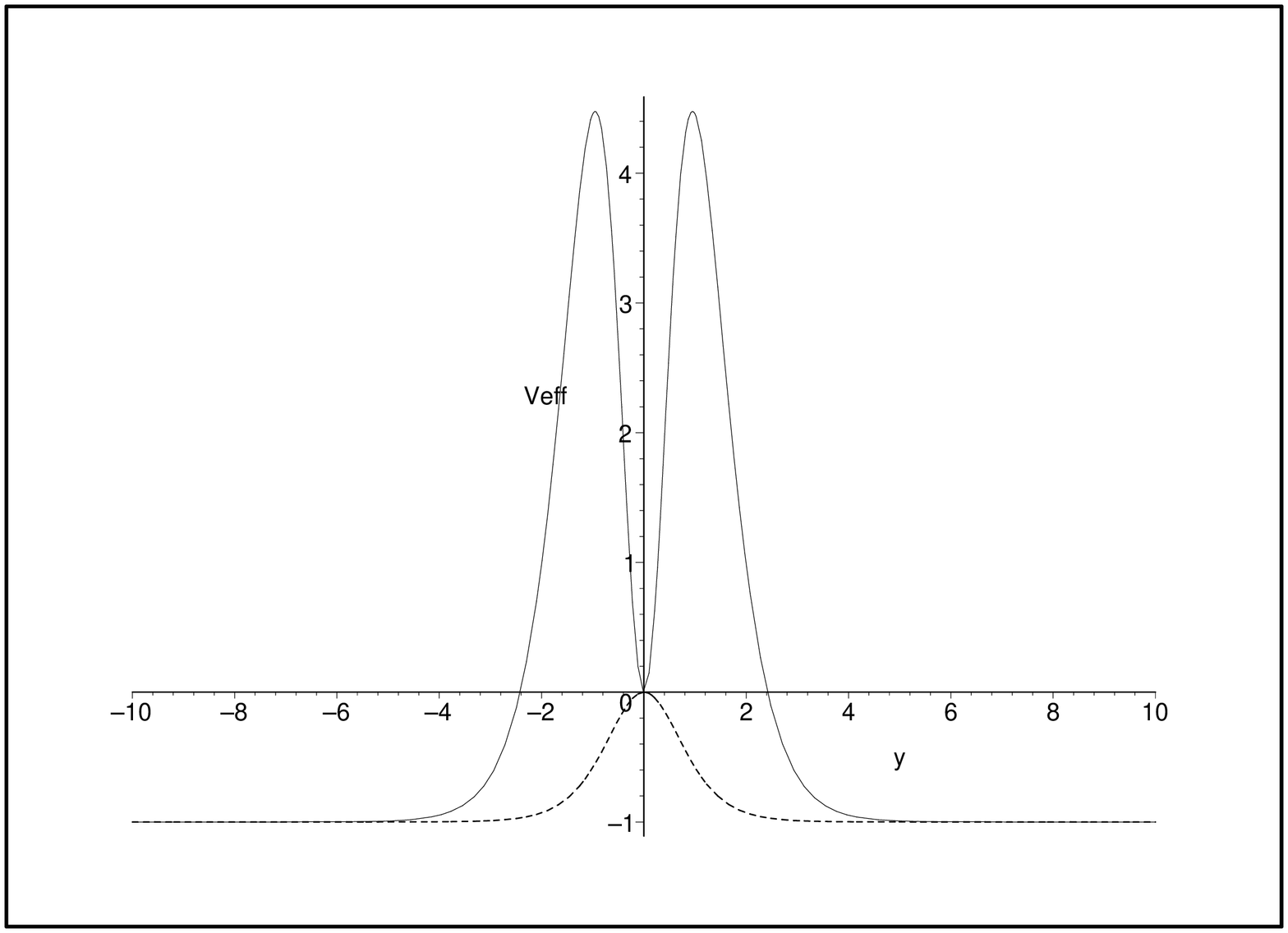}%
\caption{Confining behavior of the effective potential $V_{eff}$
when the interaction is taken into account(continuous line) for a
particular choice of the constants ($b=c=1,\kappa^{2}=2,
h=4\sqrt{2}, m=\sqrt{3}$), compared to the case when there is no
interaction(dotted line), i.e. for $h=0$.}%
\label{fig}%
\end{center}
\end{figure}

\end{document}